\documentclass[preprint2,times]{aastex62} 

\shorttitle{Temperature-dependent damping of slow waves}
\shortauthors{Krishna Prasad et al.}

\begin{document}

\title{\large The Temperature-dependent Damping of Propagating Slow Magnetoacoustic Waves} 

\correspondingauthor{S. Krishna Prasad}
\email{krishna.prasad@qub.ac.uk}

\author[0000-0002-0735-4501]{S. Krishna Prasad} 
\affiliation{Astrophysics Research Centre, School of Mathematics and Physics,
Queen's University Belfast, Belfast, BT7 1NN, UK}                              

\author{D. B. Jess} 
\affiliation{Astrophysics Research Centre, School of Mathematics and Physics,
Queen's University Belfast, Belfast, BT7 1NN, UK}                              

\affiliation{Department of Physics and Astronomy, California State University
Northridge, Northridge, CA 91330, U.S.A.}                                

\author{T. Van Doorsselaere}
\affiliation{Centre for mathematical Plasma Astrophysics, KU Leuven,
Celestijnenlaan 200B, 3001 Leuven, Belgium}                                  
 
\begin{abstract}
The rapid damping of slow magnetoacoustic waves in the solar corona has been extensively studied in previous years. Most studies suggest that thermal conduction is a dominant contributor to this damping, albeit with a few exceptions. Employing extreme-ultraviolet (EUV) imaging data from SDO/AIA, we measure the damping lengths of propagating slow magnetoacoustic waves observed in several fan-like loop structures using two independent methods. The dependence of the damping length on temperature has been studied for the first time. The results do not indicate any apparent decrease in damping length with temperature, which is in contrast to the existing viewpoint. Comparing with the corresponding theoretical values calculated from damping due to thermal conduction, it is inferred that thermal conduction is suppressed in hotter loops. An alternative interpretation that suggests thermal conduction is not the dominant damping mechanism, even for short period waves in warm active region loops, is also presented.
\end{abstract}

\keywords{magnetohydrodynamics (MHD) --- methods: observational --- Sun: atmosphere --- Sun:corona --- Sun: oscillations} 

\section{Introduction}
\label{intro}
Propagating waves along fan-like active region loops have been a common observational feature since their initial discovery in the solar corona \citep{1997ApJ...491L.111O, 1998ApJ...501L.217D, bc1999SoPh186,2000A&A...355L..23D}. Recent multi-wavelength observations have established the origin of these waves in the photosphere \citep{2012ApJ...757..160J,2015ApJ...812L..15K}, from where they could be channelled by magnetic fields into the corona \citep{2005ApJ...624L..61D,2006ESASP.624E..15E,2008ApJ...676L..85K}. It is believed that these waves are a manifestation of propagating slow magnetoacoustic oscillations that are generated via mode conversion \citep{1991LNP...388..121S, 1994ApJ...437..505C} in the lower atmospheric layers. Their physical properties found in a variety of coronal structures have been extensively studied both from theoretical and observational vantage points \citep{2009SSRv..149...65D,2011SSRv..158..397W,2012A&A...546A..50K,2016GMS...216..419B}. In the solar corona, the slow magnetoacoustic waves undergo rapid damping and are consequently visible only over a small fraction of the loop length. A number of physical mechanisms such as thermal conduction, compressive viscosity, optically thin radiation, gravitational stratification, divergence of the magnetic field, etc., are known to affect the amplitude of slow magnetoacoustic waves. Thermal conduction, however, has been put forward as the dominant contributor for their damping \citep{2003A&A...408..755D,2004A&A...415..705D}. It must be noted that slow magnetoacoustic waves also exhibit frequency-dependent damping, with stronger dissipation at higher frequencies \citep{2014ApJ...789..118K,2017ApJ...847....5K} which, as well, is shown to be consistent with generalized damping via thermal conduction \citep{2016ApJ...820...13M}.

\citet{2011ApJ...734...81M} studied slow magnetoacoustic waves propagating within a coronal loop using stereoscopic images from STEREO/EUVI \citep{2004SPIE.5171..111W} and simultaneous spectroscopic data from Hinode/EIS \citep{2007SoPh..243...19C}. It was found that thermal conduction was insufficient to explain the observed damping, and instead magnetic field divergence appeared to be the dominant factor. \citet{2011ApJ...734...81M} explained that the discrepancy was due to the relatively longer oscillation periods ($\sim10$~minutes) and colder temperatures ($\sim0.84$~MK) observed within the loop. Following a method developed by \citet{2011ApJ...727L..32V}, \citet{2015ApJ...811L..13W} estimated the polytropic index from the temperature and density perturbations corresponding to a standing slow magnetoacoustic wave observed in a hot flare loop. Based upon the value of the polytropic index ($\gamma = 1.64\pm0.08$) they obtained, \citet{2015ApJ...811L..13W} inferred that thermal conduction is suppressed and that the observed damping could be explained by a slightly enhanced compressive viscosity term, which was later validated through magnetohydrodynamic (MHD) simulations \citep{2018ApJ...860..107W}. \citet{2018ApJ...868..149K} investigated propagating slow magnetoacoustic waves in a number of active region fan-like loops and found a temperature dependency of the polytropic indices, whereby hotter loops corresponded to larger polytropic index values. However, the authors concluded that the polytropic index could be, in fact, affected by a range of physical processes, including an unknown heating mechanism, radiative losses, plasma flows, turbulence, etc., suggesting that a direct association between the polytropic index and thermal conduction cannot be unequivocally deduced. Indeed, \citet[][private communication]{Priv_Comm_Kolotkov} found that an imbalance in the embedded plasma heating and cooling processes can actually cause temperature-dependent variations in the polytropic index. Here, we study the damping of short period ($\sim$3 min) oscillations in quiescent active region fan-like loop structures, which was previously suggested to be the result of thermal conduction. The temperature dependency of the damping length is also investigated to find whether there are signatures of thermal conduction being suppressed in hotter loops. Details on the observational data used, the analysis methods employed, and the results obtained are presented in the subsequent sections, followed by a discussion of the obtained results and their implications for future studies of the solar corona.

\section{Observations}
Extreme-ultraviolet (EUV) imaging observations of solar coronal fan loops taken by the Atmospheric imaging assembly \citep[AIA;][]{2012SoPh..275...17L} on-board the Solar Dynamics Observatory \citep[SDO;][]{2012SoPh..275....3P} are utilized for the present study. AIA captures the entire Sun in 10 different wavelength channels, from which 6 are mainly dedicated to coronal observations. Using online data browsing tools, such as Helioviewer\footnote{https://helioviewer.org/}, we selected 30 different active regions (ARs) with fan-like loop structures, where propagating oscillation signatures are clearly observed. The observations of these ARs span from 2011 to 2016, although a majority of them were taken between 2012 and 2014 (i.e., during the last solar maximum). For each active region, a one-hour-long image sequence, comprising of a small subfield ($\approx$180$^{\prime\prime}\times$180$^{\prime\prime}$) surrounding the desired loop structures, is extracted for all 6 dedicated SDO/AIA coronal channels (94{\,}{\AA}, 131{\,}{\AA}, 171{\,}{\AA}, 193{\,}{\AA}, 211{\,}{\AA}, and 335{\,}{\AA}). The spatial sampling and the cadence of the data are 0.6$^{\prime\prime}$ per pixel and 12{\,}s, respectively. All of the data were processed using the \texttt{aia\_prep} routine, which is available within the Solar SoftWare (SSW) environment, to perform the roll angle and plate scale corrections required for subsequent scientific analysis. To achieve accurate alignment between the data from multiple channels, and to successfully implement some of the above processing steps, we employed the robust pipeline developed by Rob Rutten\footnote{http://www.staff.science.uu.nl/~rutte101/rridl/sdolib/}. This dataset was previously used in the study by \citet{2018ApJ...868..149K}, where complete observational details, including the locations, start times, active region numbers, etc., of the individual image sequences are listed.

\begin{figure*}[t!]
\begin{center}
\includegraphics[width=17cm]{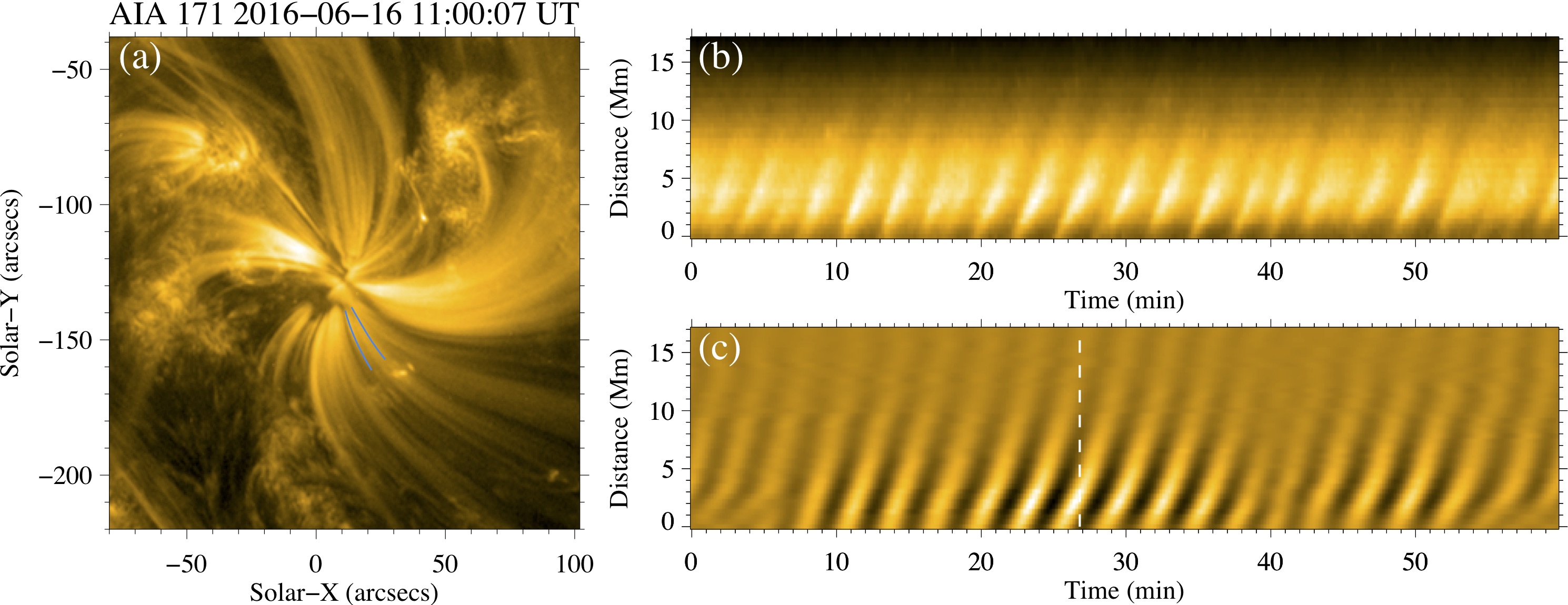}
\end{center}
\caption{(a) A snapshot of the fan-like loop structures from NOAA AR 12553 captured in the SDO/AIA 171{\,}{\AA} channel. The blue solid lines mark the boundaries of a chosen loop segment. (b) Time-distance map depicting the evolution of the loop segment shown in (a). The alternating slanted bands of brightness apparent in this map indicate the presence of propagating compressive oscillations due to slow magnetoacoustic waves. (c) Same as (b), but processed to enhance the visibility of the brightness ridges by filtering the time series at each spatial position to allow only a narrow band of frequencies around the dominant oscillation period. The white dashed line marks the temporal location chosen to study the spatial damping characteristics shown in Fig.{\,}\ref{fig2}.}\label{fig1}
\end{figure*}

\section{Analysis \& Results}
\label{sec:anres}
The fan-like loop structures within a sample active region (NOAA AR 12553) from the selected dataset are displayed in Fig.{\,}\ref{fig1}a. Compressive oscillations, with a periodicity of $\approx$180{\,}s, are found propagating outwards along these loop structures. In order to identify the oscillations and understand their propagation behavior, a time-distance map \citep{2000A&A...355L..23D} is constructed from one of the loop segments bounded by the two solid blue lines marked in Fig.{\,}\ref{fig1}a. The specific details of the method employed here are described in \citet{2012SoPh..281...67K}, but in general, the intensities corresponding to the pixels across a selected loop segment are averaged to build a one-dimensional intensity profile along the loop, with similar profiles from successive images stacked together to generate a time-distance map. The final map obtained is shown in Fig.{\,}\ref{fig1}b, with the $x$-axis displaying time in minutes and the $y$-axis displaying distance along the loop in megameters (Mm). Slanted ridges of alternating brightness, visible in this map, reveal the propagating waves along the selected loop. Previous studies of such oscillations, especially those propagating along similar fan-like loop structures that are usually rooted in sunspots, confirm their nature as propagating slow magnetoacoustic waves \citep[e.g.,][]{2012SoPh..279..427K, 2012A&A...546A..50K}. To enhance the visibility of the ridges, the time series at each spatial position was filtered to allow only a narrow band of frequencies around the dominant oscillation period to remain. The filtered time-distance map is displayed in Fig.{\,}\ref{fig1}c. It is clear that the amplitude of the oscillations is not constant, but instead varies with time and, in particular, decreases with distance along the loop from the corresponding foot point. The temporal modulation of the oscillation amplitude has been linked to the characteristics of the wave driver, with closely-spaced frequencies causing a beat-like phenomenon \citep[e.g.,][]{2015ApJ...812L..15K}, whereas the spatial damping is mainly due to physical wave dissipation and some geometrical factors. As discussed in Section{\,}\ref{intro}, thermal conduction is believed to play a key role in the observed spatial damping. 

We identify a total of 35 loop structures from the 30 active regions, where signatures of propagating slow magnetoacoustic waves are prominent. The prominence of oscillations is determined through a visual inspection of time-distance maps constructed from multiple loop structures within each active region. It may be noted that these loop structures are the same as those studied by \citet{2018ApJ...868..149K}, where the periodicity of the oscillations observed, the temperature and density of the plasma within the loop structures, the polytropic index, among other parameters, are discussed. The temperature, in particular, was derived from the peak location in the corresponding differential emission measure (DEM) curve, which was extracted by employing a regularized inversion method \citep{2012A&A...539A.146H} on the near-simultaneously observed intensities across all 6 SDO/AIA coronal channels. However, the main focus in the present study is on the damping characteristics of the oscillations. In order to study the damping properties of slow magnetoacoustic waves across the different loop structures selected, we estimate a characteristic damping length employing two independent methods, namely, a phase tracking method and an amplitude tracking method, as described in the following sections.

\subsection{Phase tracking method}
\label{sec:phase}
\begin{figure*}[t!]
\begin{center}
\includegraphics[width=17cm]{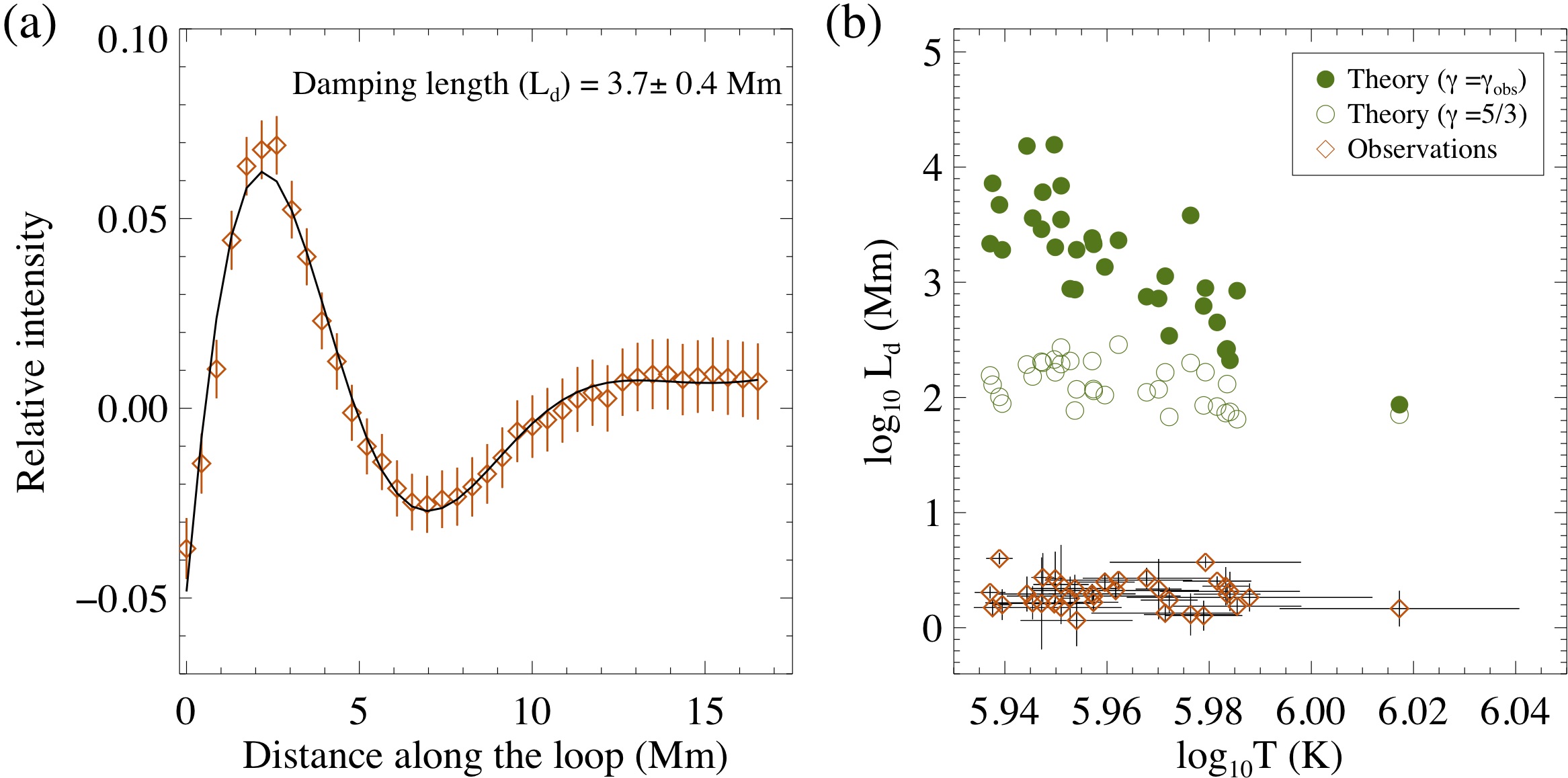}
\end{center}
\caption{(a) Spatial variation of the relative intensity at the temporal location marked by the white dashed line in Fig.{\,}\ref{fig1}c. The vertical bars denote the respective uncertainties. The solid curve represents the best fit to the data for an exponentially decaying sine wave model following Eq.{\,}\ref{eq1}. The obtained damping length value from the fitted curve is listed in the upper-right corner of the plot. (b) Damping lengths extracted from all of the selected loop structures, plotted as a function of the localized temperature on a log-log scale. The orange diamonds represent the values obtained following the phase tracking method shown in panel (a), whereas the green circles represent the theoretical values estimated from damping due to thermal conduction. The open and filled circles, respectively, highlight the values computed from a constant $\gamma$ ($=5/3$) and those computed using $\gamma$ values extracted from observations \citep{2018ApJ...868..149K}. The vertical and horizontal bars on the observed values denote the corresponding propagated uncertainties.}
\label{fig2}
\end{figure*}

A temporal location (marked by a white-dashed line in Fig.{\,}\ref{fig1}c), where the oscillation amplitude is relatively strong, is initially chosen to investigate the spatial variation of the oscillation phase. Note that the selection of this location is purely based on the strength of the oscillations as may be seen from Fig.{\,}\ref{fig1}c. Same criterion is applied to all the other loop structures studied. The filtered intensities from three consecutive frames (i.e., $\pm12$~s) around the selected temporal location are averaged to improve the signal-to-noise, then normalized by the corresponding background to construct a representative spatial intensity profile such as that shown in Fig.{\,}\ref{fig2}a. The background is constructed from the intensities obtained by smoothing the original observed values to remove any oscillations with periodicities below 10 minutes. The spatial profile clearly demonstrates a rapid decrease in the oscillation amplitude with distance along the loop. The vertical bars indicate respective uncertainties in the imaging intensities that are estimated from noise contributions in the SDO/AIA 171{\,}{\AA} channel \citep[following the methodology of][]{2012A&A...543A...9Y}, which includes noise from various sources besides the dominant photon and readout components \citep{2019ApJ...871..133J}. An exponentially decaying sine wave function of the form,
\begin{equation}
I(x)=A_{0}{\,}e^{\left(\frac{-x}{L_{d}}\right)} \mathrm{sin}\left( \frac{2 \pi x}{\lambda} + \phi \right) + B_{0}+B_{1} x \ ,
\label{eq1}
\end{equation}
is fitted to the spatial profile. Here, $I$ is the normalized pixel intensity, $x$ is the distance along the loop, $B_{0}$ and $B_{1}$ are appropriate constants, and $A_{0}$, $L_{d}$, $\lambda$ and $\phi$ are the amplitude, damping length, wavelength and phase of the oscillation, respectively. Applying the Levenberg-Marquardt least-squares minimization technique \citep{2009ASPC..411..251M}, the best fitment to the data is shown as a black solid curve in Fig.{\,}\ref{fig2}a. The damping length of the oscillation, as estimated from the fitted curve, is $3.7\pm0.4$~Mm. The orange diamond symbols in Fig.{\,}\ref{fig2}b display the damping lengths obtained from all 35 selected loop structures, plotted as a function of the corresponding localized temperature on a log-log scale. The vertical bars denote the respective uncertainties on damping length derived from the fit whereas the horizontal bars highlight the associated uncertainties on temperature propagated from the respective errors given by the regularized inversion method \citep{2012A&A...539A.146H}. Since the temperature is determined from a double-Gaussian fit to the individual DEMs \citep{2018ApJ...868..149K}, the uncertainty on peak location is estimated by scaling the corresponding error on the nearest point by a factor of $1/\sqrt N$, where, $N$ is the number of data points involved in the fit. Subsequently, to get a representative temperature value for each loop, a weighted mean across all spatio-temporal locations near the foot point is considered. The associated uncertainty is then estimated from the weighted standard deviation of values across the same locations. It may be noted that the uncertainties on loop temperature reported by \citet{2018ApJ...868..149K} are fairly small compared to those shown here (Fig.{\,}\ref{fig2}b) which is because the authors did not incorporate the temperature errors given by the DEM inversion method but simply quoted the errors obtained from the Gaussian fit alone. The actual temperature values might also marginally differ because of the weighted averages employed here. Another important aspect to note here is that in about 5 cases, the damping lengths are measured from pairs of loops from the same active region some of which exhibit distinct values. The differences in values obtained in such cases reflect the different physical conditions of the loop structures despite belonging to the same active region. 

\subsection{Amplitude tracking method}
\begin{figure*}[t!]
\begin{center}
\includegraphics[width=17cm]{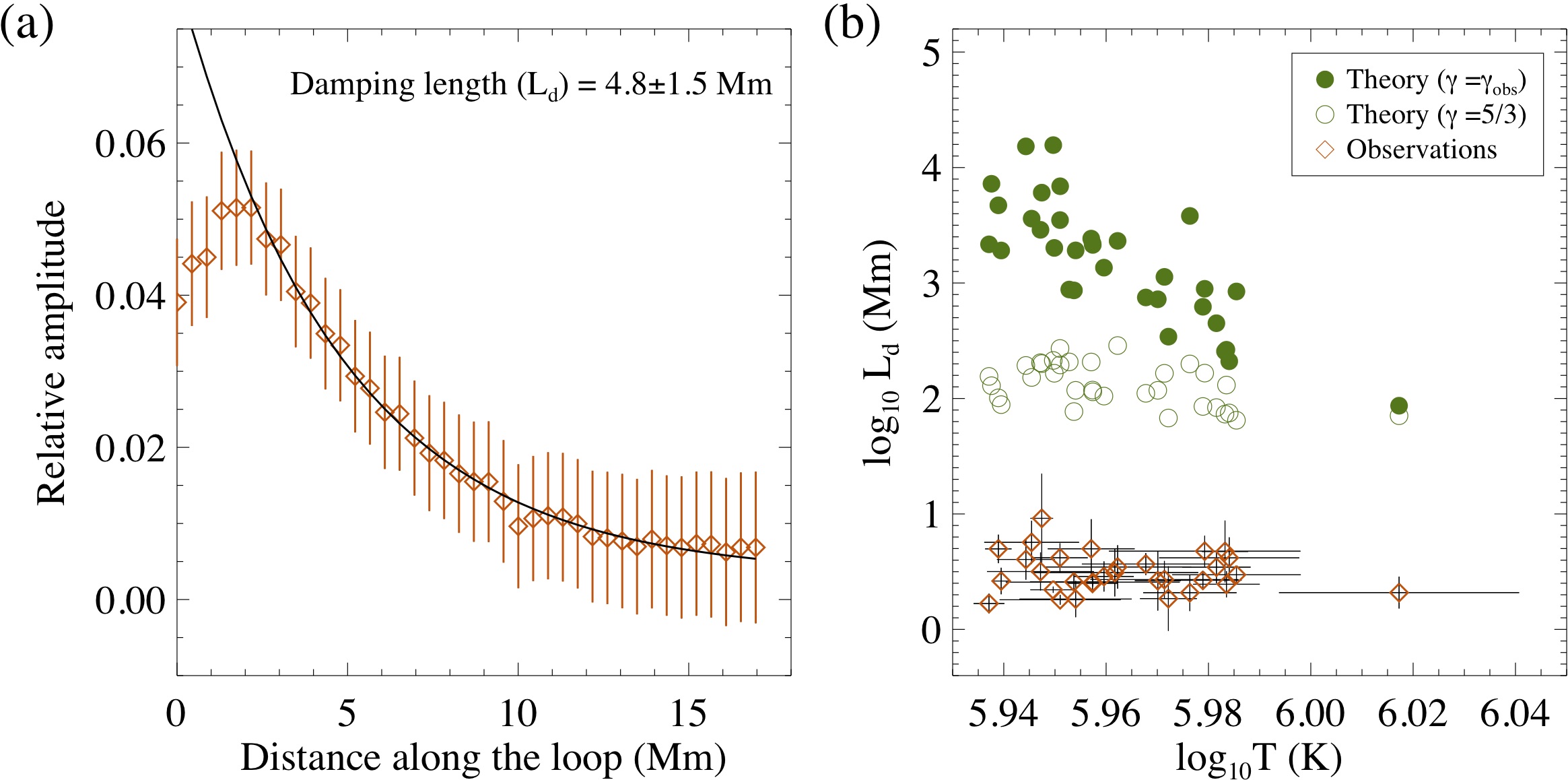}
\end{center}
\caption{(a) Relative amplitudes of the oscillations as a function of distance along the loop segment marked by the solid blue lines in Fig.{\,}\ref{fig1}a. The vertical bars denote the respective uncertainties. The black solid line represents an exponential fit to the decaying phase of the data following Eq.{\,}\ref{eq2}. The obtained damping length value from the fitted curve is listed in the plot. (b) Damping lengths extracted from all the selected loop structures plotted as a function of the local temperature on a log-log scale. The orange diamonds represent the observational values following the amplitude tracking method shown in panel (a), whereas the green circles represent the theoretical values estimated from the damping due to thermal conduction. The open and filled circles, respectively, highlight the values computed from a constant $\gamma$ ($=5/3$) and those computed using $\gamma$ values extracted from observations \citep{2018ApJ...868..149K}. The vertical and horizontal bars on the observed values denote the corresponding uncertainties.}
\label{fig3}
\end{figure*}

Since the phase tracking method involves manually choosing a specific temporal location from each of the time-distance maps, it is possible that such human intervention naturally biases the obtained results. Also, it is not trivial to apply this technique to all temporal locations since the signal-to-noise at a large number of locations is low due to aspects of amplitude modulation. To circumvent this problem and verify the reliability of our results, we estimate the damping lengths using the alternative technique of amplitude tracking. In this method, the amplitude of the oscillation, $A$, at each spatial position is directly measured in relation to the standard deviation, $\sigma$, of the respective filtered time series using $A=\sqrt{2} \sigma$. This formula assumes that the observed oscillations can be represented by a pure sinusoidal signal. The time-averaged intensities from the original time series (i.e., collapsing the time domain in Fig.{\,}\ref{fig1}b) are used as the background for normalization to obtain relative amplitudes as a function of distance along the loop. Since the amplitude at each spatial position is derived from the full time series, the median error on respective pixel intensities is used to estimate the corresponding uncertainty. The diamond symbols in Fig.{\,}\ref{fig3}a show the spatial dependence of amplitude values thus obtained for the fan-like loop structure highlighted in Fig.{\,}\ref{fig1}a. The vertical bars represent the associated uncertainties. These data were then fit with a decaying exponential model satisfying the functional form, 
\begin{equation}
A(x)=A_{0}{\,}e^{\left(\frac{-x}{L_{d}}\right)} + C \ ,
\label{eq2}
\end{equation}
where $x$ is the distance along the loop, $A_0$ and $C$ are appropriate constants, and $L_d$ is the damping length. 

The initial few locations where the wave amplitude is found to increase are ignored in order to isolate the purely decaying phase of the oscillation for fitment. The black solid line in Fig.{\,}\ref{fig3}a represents the best exponential fit obtained. The corresponding damping length is $4.8\pm1.5$~Mm, which is on the same order as that obtained using the phase tracking method outlined in Section{\,}{\ref{sec:phase}}. Following the same procedure, the damping lengths for the oscillations observed in all 35 selected loop structures have been estimated. For a handful of loop structures, it is found that the model does not converge properly, producing damping lengths either far greater than the loop length itself ($>1000$~Mm) or less than one pixel ($<0.1$~Mm). Upon inspection of the time-distance maps corresponding to these individual cases, we found that there are unusual brightenings, perhaps in the form of transient events manifesting in the loop background, at certain spatio-temporal locations, which naturally contaminate the amplitude extraction process and thereby prevent a robust fitment of the data. While this could be avoided by manually restricting the time series for each particular case, the main strength of this method was in the alleviation of human intervention. As such, we chose to ignore specific loop structures where the model fitting did not converge to commonly expected values. The orange diamond symbols in Fig.{\,}\ref{fig3}b represent the damping lengths obtained from the remaining 31 cases, plotted as a function of the loop temperature on a log-log scale. The vertical bars highlight the corresponding uncertainties on damping length whereas the horizontal bars denote the respective uncertainties on temperature. The loop temperature and the associated uncertainties are estimated in the same way as that described in Section{\,}\ref{sec:phase}.

\subsection{Theoretical calculations}
Theoretical and numerical calculations in the past have suggested that thermal conduction is the dominant physical mechanism responsible for the damping of slow magnetoacoustic waves in the solar corona \citep{2003A&A...408..755D,2004ApJ...616.1232K}. Considering one-dimensional linear wave theory for slow magnetoacoustic waves with thermal conduction as the damping mechanism \citep[e.g.,][]{2003A&A...408..755D,2012A&A...546A..50K}, the dispersion relation between the wave number, $k$, and the angular frequency, $\omega$, can be shown to be,
\begin{equation}
	d c_\mathrm{s}^4 k^4 + i \omega c_\mathrm{s}^2 k^2 - \gamma d \omega^2 c_\mathrm{s}^2 k^2 - i \omega^3=0 \ .
\label{eq3}	
\end{equation}
Here, $c_\mathrm{s}$ is the sound speed, $d=\frac{(\gamma -1)\kappa_\parallel T_0}{\gamma c^2_\mathrm{s} p_0}$ is the thermal conduction parameter, $\kappa_\parallel=\kappa_0 T_0^{5/2}$ is the parallel thermal conduction, and $p_0=2 n_0 k_B T_0$ is the gas pressure, where $\kappa_0$ is the thermal conduction coefficient, $T_0$ is the equilibrium temperature, and $n_0$ is the number density. For propagating waves, the frequency, $\omega$, is constant and in the limit of weak thermal conduction (i.e., when $d\omega \ll 1$), the solutions for wave number, $k$, may be found as $k=\frac{\omega}{c_\mathrm{s}}-i \frac{1}{L_\mathrm{d}}$, where $L_\mathrm{d}=\frac{2c_\mathrm{s}}{d\omega^2(\gamma -1)}$ is the damping length. The interested reader is referred to \citet{2016ApJ...820...13M} for a detailed derivation.

As described in Section{\,}\ref{sec:anres}, \citet{2018ApJ...868..149K} studied the same set of loop structures that are presented here. They applied a regularized inversion method \citep{2012A&A...539A.146H} on observed intensities in 6 coronal channels of SDO/AIA to compute corresponding DEM. Subsequently, by employing a double-Gaussian fit to the DEM curve, the temperature and the density of the plasma are calculated from the peak location and the area under the curve, respectively. The density is estimated by assuming the apparent width of the loop as equivalent to the emission depth along the line of sight effectively ignoring any background/foreground emission although a contribution to the latter from hot plasma is carefully discarded from the double-Gaussian fit. Using the theoretical relation between the relative oscillation amplitudes in temperature and density associated with a slow wave, the polytropic index of the plasma is determined after eliminating the corresponding phase shifts. The periodicity of the oscillations is also calculated through a simple Fourier analysis on the intensity fluctuations. Utilizing the respective values of these parameters computed by \citet{2018ApJ...868..149K} for each loop, we estimate the expected damping lengths from the above theory.

The filled green circle symbols shown in Figs.{\,}\ref{fig2}b \& \ref{fig3}b represent the theoretically computed values. It may be noted that the respective $d\omega$ values were found to reside in the interval $0.01 - 0.16$, so the assumption of weak thermal conduction (i.e., $d\omega \ll 1$) is inherently valid across our range of coronal datasets. The classical Spitzer values for thermal conduction, following $\kappa_\parallel=7.8\times 10^{-7} T_0^{5/2}$~ergs{\,}cm$^{-1}${\,}s$^{-1}${\,}K$^{-1}$, are employed in our calculations. As can be seen from the figures, the damping lengths are expected to be considerably shorter for hotter loops. Since it is very common to assume the polytropic index, $\gamma$, is equal to $5/3$ in the solar corona, we additionally compute damping lengths arising from a constant ($5/3$) value for $\gamma$. The open green circles shown in Figs.{\,}\ref{fig2}b \& \ref{fig3}b represent these values which suggest a similar but much shallower dependence on temperature. Moreover, the damping lengths in this case are shorter by up to an order of magnitude or more which clearly divulges the effect of polytropic index on the damping length. It is worth noting here that the scatter in the theoretically computed damping lengths is mainly due to the different physical conditions of the loop structures studied. 

\section{Discussion \& Concluding Remarks}
The spatial damping characteristics of propagating slow magnetoacoustic waves, observed in 35 fan-like loop structures selected from 30 different active regions, have been studied. The damping length, in particular, is measured using two independent methods: a phase tracking method and an amplitude tracking method. Employing the temperature information acquired from DEM analysis, the temperature dependence of the damping length has been investigated for the first time (Figs.{\,}\ref{fig2}b \& \ref{fig3}b). These results do not indicate any apparent decrease in damping length with temperature as would be expected by the stronger thermal conduction in that case. It may be noted that the results from previous studies \citep[e.g.,][]{2012A&A...546A..50K}, who based their conclusions on the measurement of damping lengths for a single loop structure observed in multiple temperature channels, are inconsistent with the current findings. However, those studies are purely qualitative and the results are often based on just two temperature channels. Furthermore, the sensitivities of the measured damping lengths from the intensity perturbations to the filter/instrument used \citep[e.g.,][]{2004ApJ...616.1232K} are also not taken into consideration in previous studies. In the present case, we employ damping length measurements from multiple loop structures observed in the same filter (SDO/AIA 171{\,}{\AA}). Hence, we naturally consider the current results more reliable due to the conservation of instrument characteristics across all independent measurements.

Utilizing the temperatures, densities, polytropic indices, and oscillation periods that have previously been derived for the same set of loop structures \citep{2018ApJ...868..149K}, we calculated the theoretical damping lengths expected from the dissipation due to thermal conduction. In contrast to the observations, the theoretical calculations show a steep decrease in the damping length with temperature. Damping lengths were also computed assuming a fixed value, $5/3$, for the polytropic index, in line with  the previous studies. These values display a similar but shallower dependence. The discrepancy between the observational and theoretical dependences perhaps indicates that thermal conduction is suppressed in hotter loop structures. In fact, the increase in the polytropic indices of these loops with temperature, as reported by \citet{2018ApJ...868..149K}, also implies the suppression of thermal conduction in hotter loops \citep[e.g., in accordance with][]{2015ApJ...811L..13W}, although a direct conclusion could not be drawn from these results alone since the polytropic index of the coronal plasma is dependent on several other physical processes besides thermal conduction. The current results, on the other hand, appear to show direct evidence for the suppression of thermal conduction with increasing temperature. 

Alternatively, one could argue that thermal conduction is perhaps not the dominant damping mechanism for slow magnetoacoustic waves, as previously reported by \citet{2011ApJ...734...81M} and \citet{2015ApJ...811L..13W}. Indeed, as can be seen from Figs.{\,}\ref{fig2}b \& \ref{fig3}b, the theoretical damping lengths are $2-3$ orders of magnitude higher than those obtained from the observations. One may also note that a simple visual inspection of oscillation amplitudes, in Figs.{\,}\ref{fig1}b \& \ref{fig1}c for example, reveals significant damping within 10 Mm scales whereas the expected damping lengths due to thermal conduction are at least 100 Mm or more which clearly demonstrates the extent of mismatch between the observations and the theory. The differences in the temperature dependence would further add to this discrepancy. We note, however, that the incongruity between the theory and observations is less if we consider the calculations for $\gamma=5/3$. Also, the distances measured along the observed loop structures are projected onto the image plane, meaning the obtained damping lengths are only lower limits. Nevertheless, the difference between the theoretical and the observed scales is too large to ignore, and would not likely be accounted for even if a fractional contribution from the other damping mechanisms (e.g., compressive viscosity and optically thin radiation) is included. 

Lastly, we would like to bring out some of the major caveats of our results. The temperature range of the loops investigated is limited especially considering the large uncertainties on temperature. While the magnitude of change in the expected damping lengths over the same temperature range and the extent of mismatch between the observed and theoretical values still make our results valid, it is imperative to state that a larger temperature range would make the results more reliable. Additionally, it should be noted that different DEM inversion methods can result in  different peak temperatures although the difference can be marginal depending on the temperature range investigated. Also, it can be argued whether the peak emission in a DEM sufficiently represents the plasma within the loop. Keeping these limitations in mind, we believe further investigations, both theoretical and observational, are necessary to understand the damping of slow magnetoacoustic waves in the solar corona. In particular, the impetus is on increasing the temperature range studied to include hotter loop structures to examine whether these traits are consistent across the full spectrum of coronal magnetism.

\section*{Conflict of Interest Statement}
The authors declare that the research was conducted in the absence of any commercial or financial relationships that could be construed as a potential conflict of interest.

\section*{Author Contributions}
SKP and DBJ planned and designed the study; SKP processed the data and performed the analysis; SKP wrote the first draft of the manuscript; TVD assisted in theoretical calculations; DBJ and TVD contributed to the interpretation of the results; All authors took part in the manuscript revision, read and approved the submitted version.

\section*{Funding}
The authors gratefully acknowledge the financial support from the following research grants:
\begin{itemize}
	\item STFC --- ST/K004220/1; ST/L002744/1; ST/R000891/1
	\item Randox Laboratories Ltd. --- 059RDEN-1
	\item European Research Council (ERC) --- 724326
	\item KU~Leuven --- GOA-2015-014
\end{itemize}
\section*{Acknowledgments}
The authors thank all the referees for their valuable comments/suggestions. SKP would like to thank the UK Science and Technology Facilities Council (STFC) for support. DBJ gratefully acknowledges STFC for the award of an Ernest Rutherford Fellowship, in addition to Invest NI and Randox Laboratories Ltd. for the award of a Research \& Development Grant (059RDEN-1). This project has received funding from the European Research Council (ERC) under the European Union's Horizon 2020 research and innovation programme and KU~Leuven. The SDO/AIA imaging data employed in this work are courtesy of NASA/SDO and the AIA science team. We acknowledge the use of pipeline developed by Rob Rutten to extract, process, and co-align AIA cutout data.
\section*{Data Availability Statement}
The data analyzed in this study was obtained by the Atmospheric Imaging Assembly onboard NASA's Solar Dynamics Observatory. The data are not restricted, and requests to access these datasets should be directed to the Joint Science Operations Center (\href{http://jsoc.stanford.edu/ajax/lookdata.html}{http://jsoc.stanford.edu/ajax/lookdata.html}).

\end{document}